\documentclass[11pt]{article}

\usepackage{amsmath}
\usepackage{amsfonts}
\usepackage{amssymb}
\usepackage{graphicx}
\usepackage{supertabular}
\usepackage{setspace}
\usepackage{xcolor}
\usepackage{array}
\usepackage{amsthm}
\usepackage{bm}
\usepackage{latexsym}
\usepackage{mathptmx}
\usepackage{hyperref}
\hypersetup{colorlinks=false}

\RequirePackage{cite}%
\renewcommand{\citeleft}{\bgroup\normalfont[}%
\renewcommand{\citeright}{]\egroup}%

\newcommand{\nin}{\noindent}
\newcommand{\nn}{\nonumber}
\newcommand{\be}{\begin{equation}}
\newcommand{\ee}{\end{equation}}
\newcommand{\ba}{\begin{eqnarray}}
\newcommand{\ea}{\end{eqnarray}}
\newcommand{\bal}{\begin{align}}
\newcommand{\eal}{\end{align}}

\newcommand{\dd}{{\rm d}}

\newcommand{\om}{\omega}

\newcommand{\la}{\lambda}

\newcommand{\ka}{\kappa}

\newcommand{\ga}{\gamma}
\newcommand{\ro}{\rho}

\newcommand{\ta}{\theta}
\newcommand{\Ta}{\Theta}

\newcommand{\de}{\delta}

\newcommand{\bw}{\begin{widetext}}
\newcommand{\ew}{\end{widetext}}

\def\s{\sqrt2}

\def\Q{quintessence}
\def\q{quintessence }

\def\abh{black hole }
\def\bh{black holes }
\def\aBH{black hole}
\def\BH{black holes}

\def\s{Schwarzschild }
\def\RN{Reissner-Nordstr\"om }
\def\t{thermodynamic }
\def\ts{thermodynamics }

\renewcommand{\theequation}{\arabic{section}.\arabic{equation}}

\begin{document}


\title{{\large\textbf{Thermodynamical, geometrical and Poincar\'e methods for charged \bh in presence of quintessence}}}

\author{Mustapha Azreg-A\"{\i}nou\textsuperscript{(a)}\hspace{0.009\textwidth} and
Manuel E. Rodrigues\textsuperscript{(b) (c) (d)}\thanks{E-mail: esialg@gmail.com}\\\\
{\footnotesize \textsuperscript{(a)} Ba\c{s}kent University, Department of Mathematics, Ba\u{g}l\i ca Campus, Ankara, Turkey}\\
{\footnotesize \textsuperscript{(b)} Universidade Federal do Esp\'{\i}rito Santo, Centro de Ci\^{e}ncias Exatas - Departamento de F\'{\i}sica,} \\
{\footnotesize Av. Fernando Ferrari, 514 - Campus de Goiabeiras, CEP29075-910 - Vit\'{o}ria/ES, Brazil}\\
{\footnotesize \textsuperscript{(c)} Faculdade de F\'isica, Universidade Federal do Par\'a, 66075-110, Bel\'em-Par\'a, Brazil}\\
{\footnotesize \textsuperscript{(d)} Faculdade de Ci\^{e}ncias Exatas e Tecnologia, Universidade Federal do Par\'{a},} \\
{\footnotesize Campus Universit\'ario de Abaetetuba, CEP 68440-000, Abaetetuba, Par\'{a}, Brazil}}

\date{}

\maketitle

\begin{abstract}
Properties pertaining to thermodynamical local stability of \RN \bh surrounded by \q as well as adiabatic invariance, adiabatic charging and a generalized Smarr formula are discussed. Limits for the entropy, temperature and electric potential ensuring stability of canonical ensembles are determined by the classical thermodynamical and Poincar\'e methods. By the latter approach we show that microcanonical ensembles (isolated \BH) are stable. Two geometrical approaches lead to determine the same states corresponding to second order phase transitions.

\vspace{3mm}

\nin {\footnotesize\textbf{PACS numbers:} 04.70.-s; 04.20.Jb; 04.70.Dy}

\vspace{-4mm} \nin \line(1,0){431} 
\end{abstract}

\section{\large Introduction}

Quintessence has made the subject of many papers ranging from exact solutions~\cite{C01}-\cite{cit1} to cosmological models~\cite{Diaz} (and references therein) all fueled by observational data, which were achieved through various projects~\cite{proj}, and the discovery of the acceleration of the universe. These observations lead to believe that the accelerated expansion is attributable to an exotic fluid with negative pressure making up the \Q.

The 4-dimensional spherically symmetric static solutions derived in Ref.~\cite{Kis}, with \q as source term, obey a special condition of additivity and linearity in the energy-momentum tensor. They do not exhaust the set of spherically symmetric static solutions and the quest for new solutions remains open. They have been generalized to $d$-dimensional spherically symmetric static solutions in Ref.~\cite{cit1}.

The solutions derived in Ref.~\cite{Kis} depend on four parameters: the mass and charge ($M,q$) of the \RN (RN) \abh and ($c,\om$) where $c>0$ is the \q charge, which determines the energy density $\ro_q$ of \Q, and $-1<\om <0$ is the \q state parameter.\footnote{The parameters ($c,\om$) are related by: $c\om \leq 0$. Solutions with $\om \geq 0$ also exist.} As is well known all observational cosmological data~\cite{proj} support a phenomenological equation of state $p_q=\om\ro_q$ where $p_q$ is the \q isotropic pressure. The model developed in~\cite{Kis} is anisotropic in that the local diagonal spatial components $T_i{}^j$ ($i=j$) of the energy-momentum tensor are not equal; however, their average values over spatial directions are all equal to the isotropic value $-\om\ro_q$, as it should be, leading thus to the phenomenological equation of state $p_q=\om\ro_q$.

Depending on the value of $-1<\om <0$, the solution derived in Ref.~\cite{Kis} are asymptotically flat if $-1/3\leq \om<0$ or non-asymptotically flat if $-1<\om <-1/3$. It is worth mentioning that the asymptotically flat solution~\cite{Kis}, which is a scalar field dressing an ordinary \RN \aBH, evades the ``no-scalar-hair'' theorem proven in Ref.~\cite{hair1}. The theorem is based on some symmetry settings and assumptions among which the key equality $T_t{}^t=T_{\ta}{}^{\ta}$ of the time and polar components of the energy-momentum tensor. This equality is not satisfied by the \q asymptotically flat, static, spherically symmetric \abh derived in Ref.~\cite{Kis} which is one among many other black holes bypassing the no-hair theorems~\cite{hair2,hair3}. The violation of the necessary relation $T_t{}^t=T_{\ta}{}^{\ta}$ has been noticed too for black-holes in spontaneously broken Yang-Mills gauge theories~\cite{hair3}.

To our knowledge only the non-asymptotically flat solution with $\om =-2/3$ was investigated~\cite{Kis} and its physical, geometrical and thermodynamical properties were discussed. One of the aims of this paper is to discuss those physical and thermodynamical properties of asymptotically flat solutions ($-1/3\leq \om<0$) pertaining to thermodynamic stability. Non-asymptotically flat solutions will make the subject of a subsequent work.

We shall also investigate the thermodynamic stability and phase transitions of the asymptotically flat solutions ($-1/3\leq \om<0$). To that end we shall apply the following known approaches: (1) The classical thermodynamical method, (2) the Poincar\'e method, and (3) two geometrical methods.

The classical thermodynamical method is well known in the scientific literature and has been widely applied to thermodynamic stability and phase transitions of \BH~\cite{cl1}-\cite{cit4}. As is well known, gravitating systems do not obey the linear rules for mass and entropy additions. Thus the classical method does not apply to thermodynamic ensembles of gravitating systems~\cite{cl1}-\cite{cit4} where mass or entropy is held constant (and used as a control parameter). An instance of that, the thermodynamic stability analysis or phase transitions of an isolated \abh (mass does not fluctuate) can't be carried out by either classical thermodynamical or geometrical methods which rely on the linearity hypothesis.

Poincar\'e~\cite{Poin} developed a powerful method applicable to problems pertaining to equilibrium and conditions of stability. Originally the method, known as the turning point method (TPM), was applied to the uniform rotational motion of a homogeneous liquid to determine the cases of local equilibrium and the conditions of stability of such equilibrium. Then it was applied to different situations~\cite{tr1}-\cite{bks} including problems non-tractable by classical thermodynamical or geometrical methods. It was generalized to many-parameter equilibrium families~\cite{tr6}.

The essence of the TPM method is as follows. First of all, one subdivides the space of all equilibrium configurations into 1-parameter subspaces. In each subspace, all points representing equilibrium states are related by varying one parameter called the control parameter; hence the name of linear series of equilibrium given to each subspace. The method~\cite{Poin} consists in further subdividing each linear series of equilibrium into smaller subspaces labeled stable, less stable, ..., unstable states of equilibria. The method employs the terminology of increasing or decreasing ``degree of stability'', depending on the number of negative modes of the associated Hessian matrix, and no notion of phase transition is employed. The choice of the Massieu function or the thermodynamic potential depends on the thermodynamic ensemble under consideration and in many cases there is no need to evaluate the eigenvalues of the Hessian matrix associated with the Massieu function or the thermodynamic potential in order to decide whether a linear series of equilibrium is stable~\cite{tr1}-\cite{tr5}.

Geometrical methods~\cite{quevedo1}-\cite{quevedo4} are geometric approaches which attach a measure of length to the space of all equilibrium configurations, the metrics of which are built up from a Legendre-invariant thermodynamic potential and its first and second order partial derivatives with respect to a set of extensive variables. It is well known that stability analysis results depend on the thermodynamic ensemble~\cite{cl1,cl2,cl3,tr1,tr3}, despite this fact the results derived in~\cite{quevedo1}-\cite{liu} do not make reference to any ensemble. Thermodynamic ensembles were used for the first time in~\cite{quevedo4}. The geometric methods rely on the linearity hypothesis and so are not applicable to isolated self gravitating systems.

Our interest to asymptotically flat solutions, which are not cosmologically relevant in that their scalar field does not phenomenologically represent the effects of \Q, is two fold. This will allow us to generalize the results on thermodynamic stability derived for ordinary \RN \bh~\cite{cl3}, since in this case the scalar field, dressing the \RN \aBH, does not modify the asymptotic geometry of the hole but the entropy. We emphasize that our main purpose is to investigate the thermodynamics, and not the cosmological consequences, of these \BH. On the other hand, from a theoretical point of view, it is instructive to apply the three different methods to this more involved thermodynamic problem and see how one can reach the same conclusions, albeit using different terminologies. As the classical thermodynamic method, applied to non-asymptotically flat solutions, demands different approaches~\cite{cit2,cit4,cit5}, we postpone the non-asymptotically flat case to a subsequent work.

The paper is organized as follows. In Sects.~\ref{2} and~\ref{3} we mainly discuss: horizon location, extremality conditions, relevant thermodynamic entities, generalized Smarr formula and first law of thermodynamics, adiabatic invariance and adiabatic charging of RN \bh with or without \Q. Sect.~\ref{4} is devoted to the determination of the conditions of stability of equilibrium configurations of RN \bh surrounded by \Q. We apply both, but in reversed order, the TPM and, whenever applicable, the classical thermodynamic approach. We will be able to always reach the same conclusion regarding the stability conditions but using different terminologies. In Sect.~\ref{5} we define, within the contexts of two geometric approaches, the canonical ensemble describing the thermodynamic of an RN \abh (with \Q) immersed in a heat bath then apply two geometrical methods to analyze its stability. We conclude in~Sec.~\ref{6}. An Appendix section is added to extend the analysis done in subsection~\ref{4a} to embrace all possible variations of the canonical-ensemble thermodynamic variables. We will reach the conclusion that the CE is unstable if all thermodynamic variables are allowed to vary.

\section{\large Properties of RN \bh in presence of quintessence}\label{2}

The general metric of a static and spherically symmetric spacetime is
\begin{equation}\label{2.1r}
    \dd s^2=g(r)\dd t^2-f^{-1}(r)\dd r^2-r^2\dd \Omega^2,
\end{equation}
and the general local expressions of the energy-momentum-tensor components of a static and spherically symmetric configuration are~\cite{Kis}
\begin{equation}\label{2.2r}
 T_t{}^t=A(r),\quad T_t{}^j=0,\quad T_i{}^j=C(r)r_ir^j+B(r)\de_i{}^j.
\end{equation}

To solve the field equations, $G_{\mu\nu}=\kappa T_{\mu\nu}$, for a static and spherically symmetric configuration one needs to set further conditions or ansatzes on $T_{\mu}{}^{\nu}$. If the matter source is \Q, two different ansatzes have been considered so far: (1) $C(r)=0$~\cite{C01,C02} and (2) $r_nr^nC(r)/B(r)=\text{const}$~\cite{Kis}.\footnote{In Eq.~\eqref{2.2r} we use the notation of Ref.~\cite{Kis} where it is stated that $C(r)/B(r)=\text{const}$. In fact, the correct statement is the one given here: $r_nr^nC(r)/B(r)=\text{const}$. Notation in Eq.~\eqref{2.3r} is different from that used in Ref.~\cite{Kis}.}

In this paper we are interested in the second case [$r_nr^nC(r)/B(r)=\text{const}$]. In the following we shall outline the steps leading to exact \abh solutions when \q is the matter source. Setting $A=\ro_q(r)$, $B=3\om\ga\ro_q(r)$, and $r_nr^nC(r)/B(r)=-3(1+3\ga)/\ga$, the nonvanishing components of the energy-momentum tensor take the forms
\begin{equation}\label{2.3r}
 T_t{}^t=\ro_q(r),\quad T_i{}^j=3\om\ro_q(r)\Big[-(1+3\ga)\frac{r_ir^j}{r_nr^n}+\ga\de_i{}^j\Big],
\end{equation}
where $\ga$ is a real constant. Note that the local components of $T_{\mu}{}^{\nu}$ are all different and anisotropic; however, the average values over the angles of the diagonal spatial components, $\langle T_i{}^j\rangle=-\om\ro_q\de_i{}^j$, are isotropic and do not depend on $\ga$; they are all equal and lead to the phenomenological equation of state,
\begin{equation}\label{2.4r}
    p_q=\om \ro_q,
\end{equation}
since, by definition, $\langle T_i{}^j\rangle=-p_q\de_i{}^j$ where $p_q$ is the isotropic thermodynamic pressure of the \q matter. Notice that the key equality, $T_t{}^t=T_{\ta}{}^{\ta}$, for the validity of the ``no-scalar-hair'' theorem~\cite{hair1} is violated. Even $\langle T_t{}^t\rangle=\langle T_{\ta}{}^{\ta}\rangle$ is violated unless $\om=-1$.

If the energy-momentum tensor is further endowed with the so-called ``additivity and linearity'' property, $T_t{}^t=T_r{}^r$, this fixes the value of $\ga=-(3\om+1)/6\om$ and leads to~\cite{Kis}
\begin{equation}\label{2.4r}
    g(r)=f(r).
\end{equation}
The additivity and linearity property ensures that if $T_{\mu}{}^{\nu}=\sum_{\ell}c_{\ell}(T_{\mu}{}^{\nu})_{\ell}$ is an $\ell$-term linear combination of energy-momentum tensors then the function $f-1$ is a sum of $\ell$ terms in a one-to-one correspondence with the terms of $T_{\mu}{}^{\nu}$. Restricting ourselves to the case $-1<\om<0$, the above ansatz [$r_nr^nC(r)/B(r)=\text{const}$] along with the additivity and linearity property lead to the general solution~\cite{Kis}
\begin{equation}\label{2.1}
    \dd s^2=f(r)\dd t^2-f^{-1}(r)\dd r^2-r^2\dd \Omega^2
\end{equation}
with\footnote{In the original derivation of~\eqref{2.2}, $c$ was taken negative~\cite{Kis}. We have made the substitution $-c\to2c$ for simplicity and convenience.}
\begin{equation}\label{2.2}
    f(r)=1-\frac{2M}{r}+\frac{q^2}{r^2}-\frac{2c}{r^{3\om+1}}\,,\;\, -1<\om<0 \text{ and } c>0\,.
\end{equation}
With this notation, the density of energy and isotropic pressure of \q are
\begin{equation}\label{2.3}
    \ro_q=-\frac{3\om c}{r^{3\om+3}}>0,\quad p_q=\om \ro_q<0.
\end{equation}

A first classification of the \bh described by~\eqref{2.1} and~\eqref{2.2} is based on their asymptotic behavior
\begin{align*}
&-1/3\leq \om<0: \text{ asymptotically flat solutions}\\
&-1<\om <-1/3: \text{ non-asymptotically flat solutions},
\end{align*}
and thus their physical properties depend on the sign of $3\om+1$. In this paper we shall consider the case where the asymptotic behavior of the hole is not altered by the presence of \Q, that is the case where the \abh solution is still asymptotically flat. This corresponds to $3\om+1\geq 0$ ($-1/3\leq\om<0$) with further constraints as shown below.

In the case $3\om+1=0$, the metric~\eqref{2.1} is not always asymptotically flat. If we assume $M>0$ then (1) if $1-2c>0$ ($0<c<1/2$), the metric may be brought to the following form upon performing the coordinate and parameter transformations: $t'=\sqrt{1-2c}\,t$, $r'=r/\sqrt{1-2c}$, $M'=M/(1-2c)^{3/2}$ and $q'=q/(1-2c)$:
\begin{equation}\label{2.4}
   \dd s^2=f'\dd t'^2-f'^{-1}\dd r'^2-(1-2c)r'^2\dd \Omega^2
\end{equation}
where $f'=1-2M'/r'+q'^2/r'^2$. This asymptotically flat metric~\eqref{2.4} has a conical singularity in each plane $\ta =\ta_0=\text{constant}$. The deficit angle depends on $\ta_0$ and is equal to $4\pi c$ in the plane $\ta=\pi/2$. (2) if $1-2c\leq0$, the metric~\eqref{2.1} is no longer asymptotically flat.

\begin{figure}[tbp]
\centering
  \includegraphics[width=0.33\textwidth]{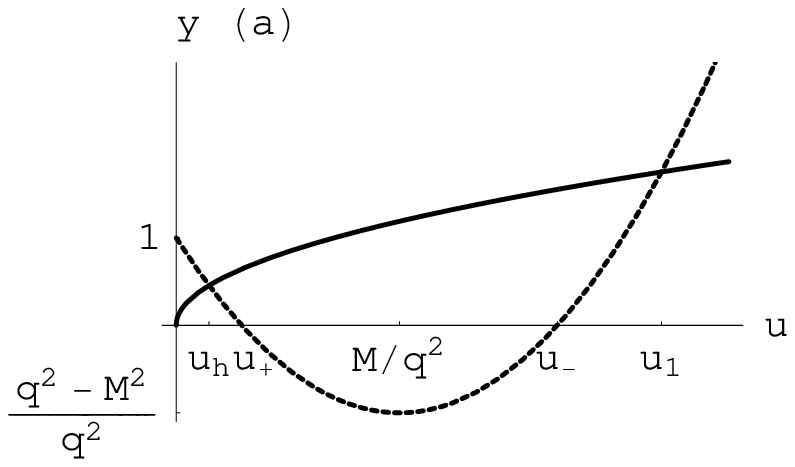}\includegraphics[width=0.33\textwidth]{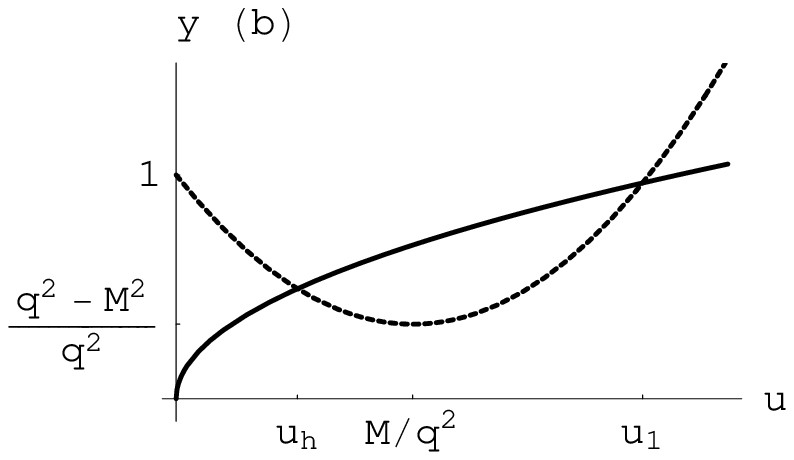}\includegraphics[width=0.33\textwidth]{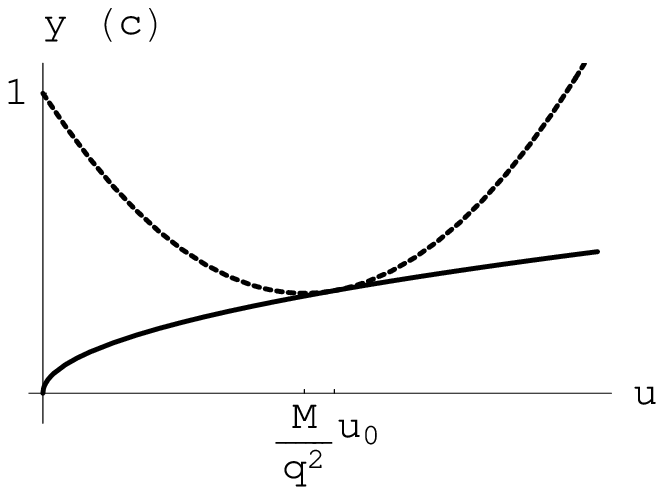}\\
  \caption{Plots of $y=1-2Mu+q^2 u^2$ (dotted line) and $y=2cu^{3\om+1}$ for $-1/3<\om<0$. (a) A \aBH: $q^2\leq M^2$. The two curves intersect at two points labeled $(u_h,u_1)$. (b) A \aBH: $q^2> M^2$ and $c>c_c$. The two curves intersect at two points labeled $(u_h,u_1)$. (c) An extreme \aBH: $q^2> M^2$ and $c=c_c$. The two curves intersect at the point $u_0$ given by~\eqref{2.9}.}\label{Fig1}
\end{figure}

Let $-1/3<\om<0$ ($1>3\om+1>0$). The horizons are defined by the condition $f(r)=0$. Setting $u=1/r$, this implies
 \begin{equation}\label{2.5}
   1-2Mu+q^2 u^2=2cu^{3\om+1}\,.
\end{equation}
The parabola $y=1-2Mu+q^2 u^2$ intersects the $u$ axis at $u_+=1/r_+$ and $u_-=1/r_-$, as shown in Fig.~\ref{Fig1} (a), with $1/u_+=r_+\equiv M+\sqrt{M^2-q^2}$ and $1/u_-=r_-\equiv M-\sqrt{M^2-q^2}$. In the absence of \Q, the right-hand side (r.h.s) of Eq.~\eqref{2.5} is zero and the remaining equation has the roots $u_+=1/r_+$ and $u_-=1/r_-$ with $r_+$ being the event horizon of an ordinary RN \aBH. The same parabola has an absolute minimum value of $(q^2-M^2)/q^2$ at $u_{\text{min}}=M/q^2$. Thus, in presence of \Q, if $q^2\leq M^2$, Eq.~\eqref{2.5} has always two roots since the graphs of $y=1-2Mu+q^2 u^2$ and $y=2cu^{3\om+1}$ always intersect, as shown in Fig.~\ref{Fig1} (a), at two points ($u_h,u_1$) such that $u_h<u_+$ and $u_1>u_-$ leading to $r_h>r_+$ where $r_h$ is the event horizon of an RN \abh surrounded by \Q. Since the entropy of a \abh is proportional to the radius of its event horizon, thus for fixed $M$ and $q$ the entropy of a \abh surrounded by \q is higher than that of an ordinary RN \aBH. This excess in entropy is attributable to the entropy of \q matter.\footnote{Since gravitating systems do not obey the linear rules for mass and entropy additions, one cannot claim that the entropy of the system (the RN \abh + \q matter) is the sum of the entropies of its components evaluated separately.}

Even in the case $q^2>M^2$, the two curves $y=1-2Mu+q^2 u^2$ and $y=2cu^{3\om+1}$ may still meet at two points ($u_h,u_1$) as shown in Fig.~\ref{Fig1} (b). Hence, the presence of \q makes it possible for the \abh to absorb more charge than the ordinary RN \abh before it becomes a naked singularity. For fixed $\om$, $M$ and $q^2/M^2>1$ there is a critical value $c_c$ of $c$ below which the two curves do not intersect. $c_c$ is such that the two curves $y=1-2Mu+q^2 u^2$ and $y=2cu^{3\om+1}$ have a common tangent line at the unique point of intersection $u_0>0$:
\begin{align}
\label{2.6}&1-2Mu_0+q^2 u_0{}^2=2c_cu_0{}^{3\om+1}\\
\label{2.7}&-M+q^2u_0=c_c(3\om+1)u_0{}^{3\om}\,.
\end{align}
Eliminating $c_cu_0{}^{3\om}$ leads to
\begin{equation}\label{2.8}
    (1-3\om)q^2u_0{}^2+6M\om u_0-(3\om+1)=0
\end{equation}
which has only one positive root given by
\begin{equation}\label{2.9}
    u_0=\frac{\sqrt{9\om^2(M^2-q^2)+q^2}-3\om M}{(1-3\om)q^2}=\frac{\sqrt{9\om^2M^2+(1-9\om^2)q^2}-3\om M}{(1-3\om)q^2}\,.
\end{equation}
Using this in~\eqref{2.7} we obtain
\begin{equation}\label{2.10}
    c_c=\frac{q^2u_0-M}{(3\om+1)u_0{}^{3\om}}
\end{equation}
for an extreme \abh solution. Since the r.h.s of~\eqref{2.7} is positive, we must have $u_0>M/q^2$, which is directly derived from~\eqref{2.9}. The roots ($u_h,u_1$) are such that
\begin{equation}\label{2.11}
    u_h< u_0< u_1\text{ and } M/q^2<u_1\quad (\text{for }c>c_c)
\end{equation}
but $u_h$ may be greater or smaller than $M/q^2$.

For fixed ($\om, M,q^2/M^2>1$) we conclude that if $c< c_c$, the metric~\eqref{2.1} is a naked singularity and if $c>c_c$, the solution is a \abh with an event horizon at $r_h$ and an inner horizon at $r_1$. The case $c=c_c$ corresponds to an extreme \abh with an event horizon at $r_0=1/u_0$ with $u_0$ given by~\eqref{2.9}.

We would formulate the last statement in terms of fixed ($\om, M,c$) if it were possible to solve~\eqref{2.10} for $q^2/M^2$. An extreme \abh is obtained (1) upon fixing ($\om, M,q^2/M^2>1$) and decreasing the value of $c$ till $c=c_c$, as done in Fig.~\ref{Fig1}, or (2) upon fixing ($\om, M,c$) and increasing $q^2/M^2$ till $q^2/M^2=(q^2/M^2)_{c}$. The case (2) is generally not possible because we cannot invert~\eqref{2.10}, however, for fixed $c\ll 1$ and given ($\om,M$) the extreme RN \abh in presence of \q is approximated by
\begin{equation}\label{2.12}
    \Big(\frac{q^2}{M^2}\Big)_c= 1+\frac{c}{M^{3\om+1}}+\mathcal{O}\Big(\frac{c}{M^{3\om+1}}\Big)^2\qquad (c\ll 1)\,.
\end{equation}
Recall that $1>3\om+1>0$, so that the contribution of the second term is not neglected even for relatively large \BH. For $3\om+1=0$ ($\om =-1/3$) and $0<c<1/2$, it is easy to show using~\eqref{2.4} or~(\ref{2.1}, \ref{2.2}) that the solution is a \abh if $q^2/M^2<1/(1-2c)$ and that the extreme \abh corresponds to
\begin{equation}\label{2.13}
    \Big(\frac{q^2}{M^2}\Big)_c=\frac{1}{1-2c}\quad (0<1-2c<1)\,.
\end{equation}

\section{\large Thermodynamic of RN \bh with quintessence}\label{3}

In this section we only consider asymptotically flat \abh solutions which correspond to
\begin{align}
\label{3.0}&[(q^2\leq M^2 \text{ and any } c>0) \text{ or } (q^2> M^2 \text{ and } c>c_c)] \text{ if } -1/3<\om<0\\
\label{3.00}&\Big[\Big(q^2\leq M^2 \text{ and } \frac{1}{2}>c>0\Big) \text{ or } \Big(q^2> M^2 \text{ and }  \frac{1}{2}>c>\frac{q^2-M^2}{2q^2}\Big)\Big] \text{ if } \om =-1/3\,.
\end{align}

Our thermodynamic system is (the RN \abh + \q matter) the metric of which is given by~\eqref{2.1} and~\eqref{2.2}. Consequently, any parameter, which expresses the system's solution~\eqref{2.1} and~\eqref{2.2}, is an appropriate thermodynamic state variable for the description of the thermodynamics and phase transitions of the system. We may take as thermodynamic state variables of these \bh the entropy $S$ that is, as we emphasized in the previous section, the total entropy of our system including that of \q matter, the electric charge $q$, and the \q charge $c$. This very choice of thermodynamic state variables was made in~\cite{cit1} for the uncharged $d$-dimensional spherically symmetric static solutions with \Q. As a general approach, every parameter that enters the metric functions may be chosen as a thermodynamic state variable, as this is done for the cosmological constant too~\cite{cosmo1,cosmo2}.

We first consider the case $-1/3<\om<0$. The total entropy of such \bh surrounded by \q is $S=\pi r_h{}^2$, which is the area of the horizon by $4$; however, for simplicity of notation we will work with the quantity $s=S/\pi$.  Using $f(r_h)=0$, we express the (gravitational) mass $M$ in terms of ($s,q,c$) as follows
\begin{equation}\label{3.1}
    M=\frac{s+q^2-2cs^{W_-}}{2\sqrt{s}}\qquad \big(W_{\pm}=\frac{1}{2}\pm \frac{3\om}{2}\big)\,.
\end{equation}
It is straightforward to show that, in presence of \Q, $M$ is still the (internal) energy of the \aBH. For that purpose we need to show that $(\partial M/\partial s)_{q,c}$ is indeed the temperature of the \aBH. The latter is proportional to the surface gravity $T=\ka/2$ and $\ka = \partial_{r}f(r)/2$. Using these equations we express $T$ is terms of the entropy and mass as
\begin{equation}\label{3.2}
    T=\frac{1}{2}\Big[\frac{M}{s}-\frac{q^2}{s^{3/2}}+\frac{(3\om+1)c}{s^{(3\om+2)/2}}\Big].
\end{equation}
Replacing $M$ by the r.h.s of~\eqref{3.1} reduces the r.h.s of~\eqref{3.2} to $(\partial M/\partial s)_{q,c}$.

Let ($A_0,A_Q$) be the functions
\begin{equation*}
    A_0=\Big(\frac{\partial M}{\partial q}\Big)_{S,c}\,,\;A_Q=\Big(\frac{\partial M}{\partial c}\Big)_{S,q}\;\text{ along with }\;T=\Big(\frac{\partial M}{\partial s}\Big)_{q,c}\,.
\end{equation*}
Using~\eqref{3.1}, we obtain
\begin{align}
\label{3.3}&A_0=\frac{q}{\sqrt{s}}\,,\quad A_Q=-\frac{1}{s^{3\om/2}}\,,\\
\label{3.4}&T=\frac{s-q^2+6c\om s^{W_-}}{4s^{3/2}}\,.
\end{align}
Using $1/s=u_h{}^2$ in~\eqref{3.2}, we see that $T\propto [c(3\om+1)u_h{}^{3\om}+M-q^2u_h]$. For the extreme \aBH, $c=c_c$ and $u_h=u_0$, so by~\eqref{2.7}, $T=0$.

Note that the r.h.s of~\eqref{3.1} is a homogenous function of ($s,q^2,c^{2/(3\om+1)}$) of degree $1/2$:
\begin{equation*}
    M(\la s,\la q^2,\la c^{2/(3\om+1)})=\la^{1/2}M(s,q^2,c^{2/(3\om+1)})\,.
\end{equation*}
By Euler's theorem we obtain the generalized Smarr formula
\begin{equation}\label{3.5}
    M=2Ts+A_0q+(3\om+1)A_Q c
\end{equation}
where $A_0$ is the electric potential and $A_Q$ is the potential associated with \Q. It is straightforward to check that changes in the thermodynamic state variables ($s,q,c$) by amounts ($\dd s,\dd q,\dd c$) result in
\begin{equation}\label{3.6}
    \dd M=T\dd s+A_0\dd q+A_Q\dd c
\end{equation}
which is the first law of black-hole thermodynamics for RN \bh in a background of \Q. The right-hand side of~\eqref{3.6} states that any change in the radius of the event horizon, in the electric charge of the \abh or in the density of \q will affect the mass parameter of the \abh by the amount given in the right-hand side. Thus, changes in the density of \q affects the internal energy of the \abh and this fact justifies the choice of $c$ as a thermodynamic variable~\cite{cit1}.

For the case $\om=-1/3$ we use~\eqref{2.4} where the hole has mass $M'$ and charge $q'$. The horizon area is multiplied by $1-2c$ so that the entropy divided by $\pi$ is $s'=(1-2c)(M'+\sqrt{M'^2-q'^2})^2$. Let $s\equiv s'/(1-2c)=(M'+\sqrt{M'^2-q'^2})^2$, then
\begin{equation}\label{3.6a}
    M'=\frac{s+q'^2}{2\sqrt{s}}\;\text{ and }\;T'=\frac{s-q'^2}{4 (1-2c)s^{3/2}}
\end{equation}
[where $T'\equiv (\partial M'/\partial s')_{q',c}$]. With $A'_0=(\partial M'/\partial q')_{S',c}$, the generalized Smarr formula for this case takes the form
\begin{equation}\label{3.6b}
    M'=2T's'+A'_0q'\,.
\end{equation}

\paragraph{Particle absorption--emission: Adiabatic invariance.}

Without loss of generality, we assume $q>0$. As shown in~\cite{bek}, a particle of mass $m$ and electric charge $\epsilon >0$ moving in the geometry described by~(\ref{2.1}, \ref{2.2}) has the conserved energy for radial motion
\begin{equation}\label{3.7}
    E=m\sqrt{f(r)+(\dd r/\dd \tau)^2}+\frac{\epsilon q}{r}
\end{equation}
where $\tau$ is the proper time. If the particle crosses the horizon $r_h$, this incurs changes in ($M,q$): $\dd M=E$, $\dd q=\epsilon$, which are related by~\eqref{3.6} [here we assume that the motion of the particle does not affect the density of \Q, that is we take $\dd c=0$]
\begin{equation}\label{3.8}
    2\Big(\dd M-\frac{q\dd q}{r_h}\Big)=[1-q^2u_h{}^2+6c\om u_h{}^{3\om+1}]\dd r_h\,.
\end{equation}
Using $f(r_h)=0$, we bring it to the form
\begin{equation}\label{3.9}
    \dd M-\frac{q\dd q}{r_h}=-[W_- q^2u_h{}^2+3M\om u_h-W_+]\dd r_h
\end{equation}
where the expression inside the square parentheses is proportional to the left-hand side (l.h.s) of~\eqref{2.8}. Since, by~\eqref{2.11}, $u_h<u_0$, the coefficient of $\dd r_h$ in~\eqref{3.9} is positive. Thus
\begin{equation}\label{3.10}
    {\rm sgn}\Big(\dd M-\frac{q\dd q}{r_h}\Big)= {\rm sgn}\,(\dd r_h)\geq 0\;\text{ (by the second law)}
\end{equation}
which we rewrite for the charged particle
\begin{equation}\label{3.11}
    {\rm sgn}\Big(E-\frac{q\epsilon}{r_h}\Big)= {\rm sgn}\,(\dd r_h)\geq 0\;\text{ (by the second law)}\,.
\end{equation}
Hence, if the particle's energy is $E_0\equiv q\epsilon /r_h$, the latter reaches the horizon with a zero speed: $\dd r/\dd \tau=0$ (as seen from~\eqref{3.7}). In this case the particle is adiabatically accreted by the hole causing no change in the horizon's area: $\dd r_h=0$ [by~\eqref{3.11}].

If $E>E_0$ the particle plunges into the hole with some kinetic energy causing the horizon to expand by the amount
\begin{equation*}
    (E-E_0)/[W_+ -3M\om u_h - W_- q^2u_h{}^2]
\end{equation*}
(particles with energy $E<E_0$ cannot reach the horizon).

Conversely, as shown in~\cite{G}, \bh with relatively high temperature radiate electrically charged particles in the superradiant regime. The strong gravitational field near the horizon creates two particles of opposite charges. The total energy lost by the field is greater than the energy carried by each particles and in any case\footnote{Excluded is the case where both particles fall into the hole. If we neglect radiations par the particles, this corresponds to $\dd M=0$, $\dd q=0$ and $\dd r_h=0$.} $\dd M<0$. If $\dd q>0$, that is the particle with positive charge falls into the hole and the particle with negative charge escapes to infinity, the l.h.s of~\eqref{3.10} is negative (since $\dd M<0$) leading to $\dd r_h<0$ which is not favored by the second law. Thus, by the second law of thermodynamics, the \abh may radiate charges of only the same sign as its own charge $q$ reducing its charge upon receiving electric charges of opposite sign to its charge so that $\dd q<0$. The process may continue till the entropy reaches its maximum values or proceed adiabatically (i.e. the charge $\epsilon$ of the falling particle is such that $\dd M=q\epsilon/r_h<0$) till the temperature drops.

\paragraph{Adiabatic charging of an RN \aBH.}

We consider an ordinary RN \abh (BHn), where $c=0$, along with another RN \abh surrounded by \q (BHq) for given and fixed values of ($\om,c$) with $-1/3<\om<0$. Each hole has mass $M_0$ and charge $q_0$ satisfying $M_0/q_0>1$ as in Fig.~\ref{Fig1} (a). The horizons, $r_+$ of (BHn) and $r_h$ of (BHq), are such that $u_h<u_+$.

Let $r_e$ denotes $r_+$ or $r_h$ ($u_e=1/r_e$ denotes $u_+=1/r_+$ or $u_h=1/r_h$). If the two \bh are charged adiabatically, $r_e$ remains constant (throughout the rest of this section $u_+$ and $u_h$ are then taken as constants), so that by~\eqref{3.8}, $\dd M=q\dd q/r_e$ leading to
\begin{equation}\label{3.12}
    M=\frac{u_e}{2}q^2+D\qquad (D\equiv M_0-\frac{u_e}{2}q_0{}^2)
\end{equation}
where the constant $D>0$ since $u_e<M_0/q_0{}^2$ as shown in Fig.~\ref{Fig1} (a). Using~\eqref{2.5} it is easy to show
\begin{align}
\label{3.13}&\text{For (BHn): }2Du_+=1\qquad (2D=r_+)\\
\label{3.14}&\text{For (BHq): }2Du_h=1-2cu_h{}^{3\om+1}<1\,.
\end{align}

While the two holes are being charged (at the same rate), the point $M/q^2$ [Fig.~\ref{Fig1} (a)] moves progressively to the left since
\begin{equation*}
    \frac{M}{q^2}=\frac{u_e}{2}+\frac{D}{q^2}
\end{equation*}
is obviously a decreasing function of $q^2$. As $M/q^2$ meets first $u_+$, (BHn) turns into an extreme \abh and the charging process ends [since~\eqref{3.10} is no longer valid] by cumulating the total charge and mass ($q_t,M_t$):
\begin{equation}\label{3.15}
    \text{For (BHn): }u_+=\frac{u_+}{2}+\frac{D}{q_t{}^2}\Rightarrow q_t=r_+\,,\;M_t=r_+
\end{equation}
as expected (the subscript ``$t$'' for total). What happens at this moment to (BHq)? Let us look at the derivative of $M/q=u_eq/2+D/q$:
\begin{equation*}
    \frac{\partial(M/q)}{\partial q}=\frac{u_e}{2}-\frac{D}{q^2}\,.
\end{equation*}
This is zero if $q^2=2D/u_e$. For (BHn), using~\eqref{3.13} along with $u_e=u_+$, we obtain  $q=r_+=q_t$, which means that $M/q$ reaches its minimum value, 1, at the end of the charging process. For (BHq) the situation is quit different: At the moment $M/q^2$ meets $u_+$, the function $M/q$ is still decreasing, so that (BHq) cumulates more charge and mass than (BHn), and reaches its minimum value at the moment $M/q^2$ meets $u_h$ with
\begin{equation}\label{3.16}
    \text{For (BHq): }(M/q)_{\text{min}}=\sqrt{2Du_h}<1.
\end{equation}
As the point $M/q^2$ passes $u_h$, the ratio $M/q$ starts to increase but remains smaller than 1. The charging process ends when (BHq) turns into an extreme \abh at the moment the point $u_0$ [Fig.~\ref{Fig1} (c)] meets $u_h$. The total charge and mass of (BHq) are then given by [use~\eqref{2.8} and~\eqref{3.12}]
\begin{equation*}
     (1-3\om)q_t{}^2u_h{}^2+6M_t\om u_h-(3\om+1)=0 \text{ and } 2M_t=u_hq_t{}^2+2D
\end{equation*}
leading to
\begin{align}
\label{3.17}&\text{For (BHq): }q_t=r_h\sqrt{1+3\om(1-Du_h)}<r_h\\
\label{3.18}&\text{For (BHq): }2M_t=r_h[(1-3\om)2Du_h+(3\om+1)]<2r_h\,.
\end{align}
It is easy to check that:
\begin{equation}\label{3.19}
 \text{For (BHq): }M_t/q_t<1\,,r_+<q_t<r_h \text{ and } r_+<M_t<r_h \,.
\end{equation}

\section{\large Thermodynamic local stability}\label{4}

In this section we only consider thermodynamic processes for which $c$ is held constant. The second law of (\aBH) \ts governs all the criteria for \t stability. These criteria, in form, depend on how we consider the \t system (the RN \abh + \q matter). For short, the system will be called: ``\aBH.'' We shall consider canonical and microcanonical ensembles (CE, ME, respectively). The CE will consist of the \abh in equilibrium with its thermal radiation, treated as a reservoir (heat bath) at constant temperature and the ME will be the case of a \abh isolated from its surroundings. It has been shown in many applications that the thermodynamic local stability of CEs may equally be treated by classical thermodynamic approaches based on the Hessian matrix of the entropy or, equivalently, of the energy~\cite{cl1,cl2,cl3} or by the TPM~\cite{tr1} derived by Poincar\'e~\cite{Poin}. Because of the non-additivity of entropy and mass in general relativity~\cite{tr2}, the classical thermodynamic approach does not apply to isolated \BH, for which we shall then apply the TPM.

\subsection{CE: Classical thermodynamic approach}\label{4a}

We assume that the hole is immersed in a thermal bath at constant temperature $T$. Applying the classical thermodynamic approach, we denote by $\dd s_b$ the change in the entropy of the bath. Then any possible change in the state of the system requires $\dd s+\dd s_b\geq 0$. Conversely, if fluctuations try to take the system out of equilibrium with the reservoir, that is if
\begin{equation}\label{4.3}
    \dd s+\dd s_b< 0
\end{equation}
for allowed changes in the system's and reservoir's parameters, the system cannot leave the current state which is said to be in stable equilibrium with the bath. The inequality~\eqref{4.3} is the condition from which all criteria for local stability of systems in contact with reservoirs are derived. Since any \t process is considered reversible for the (huge) reservoir, from~\eqref{3.6} it follows that $\dd s_b=(\dd M_b-A_0\dd q_b)/T$ ($\dd c\equiv 0$). Exchanges between the system and the reservoir obey the conservation rules: $\dd M_b=-\dd M$ and $\dd q_b=-\dd q$. Using these equations in~\eqref{4.3} leads to
\begin{equation}\label{4.2}
    \dd M>T\dd s+A_0\dd q\,.
\end{equation}
In~\eqref{4.2} we are using the mass-energy as a fundamental thermodynamic quantity, instead of the entropy, because it is not possible to reverse~\eqref{3.1} and express $s$ in terms of $M$.

By the first law, $\dd M=T\dd s+A_0\dd q$, the condition~\eqref{4.2} is not sensitive to first order changes in the allowed parameters. Keeping up to second partial derivatives of $M$ with respect to the extensive parameters ($s,q$) in the Taylor expansion of $M$, we obtain the equivalent condition (which may be written as a Hessian matrix, the general case of which, where $\dd c\neq0$, is discussed in the Appendix and leads to the instability of the CE):
\begin{equation}\label{4.4}
    \frac{1}{2}\Big(\frac{\partial^2 M}{\partial s^2}\Big)_{q,c}\dd s^2+\Big(\frac{\partial^2 M}{\partial s \partial q}\Big)_{c}\dd s\dd q+
    \frac{1}{2}\Big(\frac{\partial^2 M}{\partial q^2}\Big)_{s,c}\dd q^2 >0
\end{equation}
where all first order terms canceled out. With
\begin{equation}\label{4.5}
    \Big(\frac{\partial^2 M}{\partial s^2}\Big)_{q,c}=\frac{T}{C_q}\,,\;\Big(\frac{\partial^2 M}{\partial s \partial q}\Big)_{c}=\Big(\frac{\partial T}{\partial q}\Big)_{s,c}\,,\;\Big(\frac{\partial^2 M}{\partial q^2}\Big)_{s,c}=\Big(\frac{\partial A_0}{\partial q}\Big)_{s,c}
\end{equation}
[$T$ and $A_0$ have been defined in the equation preceding~\eqref{3.3} and their expressions are given in~(\ref{3.3}, \ref{3.5})], the condition~\eqref{4.4} implies\footnote{We would like to show the analogy with classical \ts via the following correspondences where ($P,V$) are the pressure and volume of the classical system: $\dd q\to -\dd V$, $A_0\to P$ and $C_q\to C_V$. In a similar way, $(\partial q/\partial A_0)_{s,c}/q$, which may be called the factor of adiabatic charge, corresponds to the adiabatic compressibility $-(\partial V/\partial P)_{S}/V$. Similar terminology has been used in~\cite{Ch}.}
\begin{equation}\label{4.6}
    C_q>0\,,\;\Big(\frac{\partial A_0}{\partial q}\Big)_{s,c}>0 \,\text{ and }\, \Big(\frac{\partial T}{\partial q}\Big)_{s,c}^2<\frac{T}{C_q}\Big(\frac{\partial A_0}{\partial q}\Big)_{s,c}\,.
\end{equation}
Here $C_q$ is the specific heat at constant charge
\begin{align}
C_q=&\Big(\frac{\partial M}{\partial T}\Big)_{q,c}=T/\Big(\frac{\partial^2 M}{\partial s^2}\Big)_{q,c}\nn\\
\label{4.7}\quad =&\frac{2s(s- q^2+6c \omega  s^{W_-})}{3q^2-s-6c\omega (2+3\omega )s^{W_-}}\,.
\end{align}
Note that if $T$ is constant (the case where the hole is immersed in a reservoir), the l.h.s of the third inequality in~\eqref{4.6} is zero.

The mass $M$, given by~\eqref{3.1}, is supposed to be positive, so we have the extra condition
\begin{equation}\label{4.8}
    2\sqrt{s}M=s+q^2-2cs^{W_-}>0\,.
\end{equation}

Usually, in classical thermodynamics, all relevant thermodynamic quantities pertaining to the reservoir ($M_b,q_b,\cdots$), but the temperature, are allowed to fluctuate. In this paper we shall allow $T $ to fluctuate too and investigate separately the cases $T$ constant and $T$ fluctuating.

\subsubsection{\pmb{$T$} constant}\label{4a11}

Condition~\eqref{4.4} or~\eqref{4.6} reduces to $C_q>0$ and $T>0$. Since $s>0$ and $C_q\propto T$, we have to solve simultaneously
\begin{equation}\label{4.9}
    \text{(a): }s- q^2+6c \omega  s^{W_-}>0 \,\text{ and }\,\text{(b): }3q^2-s-6c\omega (2+3\omega )s^{W_-}>0\,.
\end{equation}

For $c=0$ we recover the conditions for ordinary RN \BH:
\begin{equation}\label{4.10}
    q^2<s<3q^2
\end{equation}
with $s=(M+\sqrt{M^2-q^2})^2$ leads to the known conditions for local stability~\cite{cl3}
\begin{equation}\label{4.11}
    3/4<q^2/M^2<1\,.
\end{equation}

Now back to \q case $c>0$. For $-1/3<\om<0$ we have $1/2<W_-<1$ and $1<2+3\om<2$. In order to solve the inequality~\eqref{4.9} (a), we consider the line $y(s)=s- q^2$ and the concave-down curve $y(s)=-6c \omega  s^{W_-}$ (the graph of which is similar to that of $y=\sqrt{s}$), it is easy to see that they do intersect at one and only one point $s_1(\om)>q^2$. Thus $T>0$ if $s>s_1(\om)$, which solves~\eqref{4.9} (a). Similarly the line $y(s)=s-3 q^2$ and the concave-down curve $y(s)=-6c \omega  (2+3\om)s^{W_-}$ intersect at one and only one point $s_2(\om)>3q^2$. Thus~\eqref{4.9} (b) is solved by $s<s_2(\om)$. The \abh is locally stable if
\begin{equation}\label{4.12}
    s_1(\om)<s<s_2(\om)\;\text{ and }\;M>0\,.
\end{equation}
The first condition in~\eqref{4.12} is a generalization of~\eqref{4.10} where $q^2$ and $3q^2$ have been shifted to the right [$s_1(\om)>q^2$, $s_2(\om)>3q^2$]. The problem, however, cannot be solved as in~\eqref{4.11} since ($s_1,s,s_2$) are not known explicitly in terms of ($M,q,\om,c$).

In the case of ordinary RN \bh ($c=0$), the mass, $2\sqrt{s}M=(s+q^2)$, is a positive function of the entropy. In the case of RN \bh surrounded by \Q, this is not guaranteed a priori and we need the solve the extra inequality~\eqref{4.8}. Consider the line $y(s)=s+ q^2$ and the concave-down curve $y(s)=2c  s^{W_-}$. For fixed ($\om,c$), if we choose $q^2>q_c{}^2\equiv W_+(2cW_-)^{1/W_+}/W_-$ [for ordinary RN \bh ($c=0$), $q_c{}^2=0$], the line and the curve do not intersect and $M>0$ for all $s$ and if  $q^2\leq q_c{}^2$, they meet at two points $s_4(\om)\leq s_3(\om)$ where $M>0$ for $s>s_3(\om)$ or $s<s_4(\om)$.

\begin{figure}[tbp]
\centering
  \includegraphics[width=0.47\textwidth]{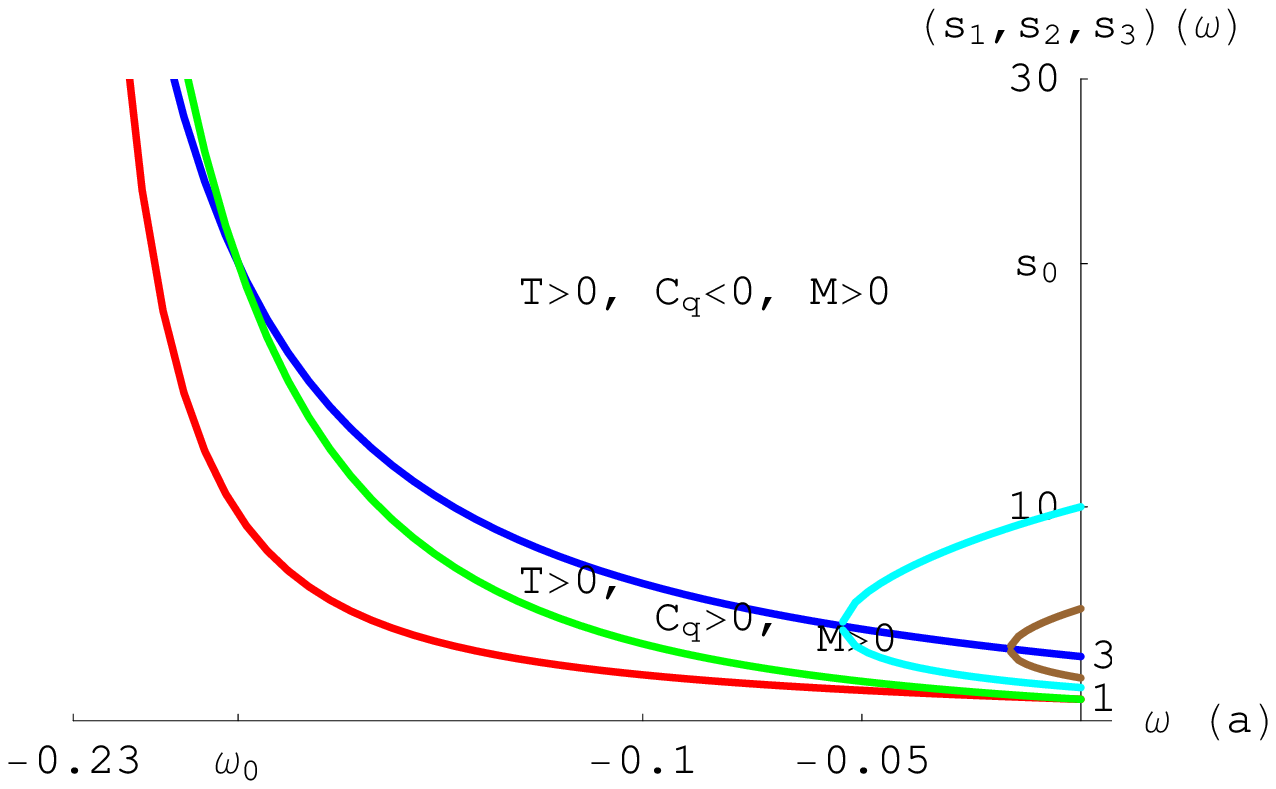} \includegraphics[width=0.47\textwidth]{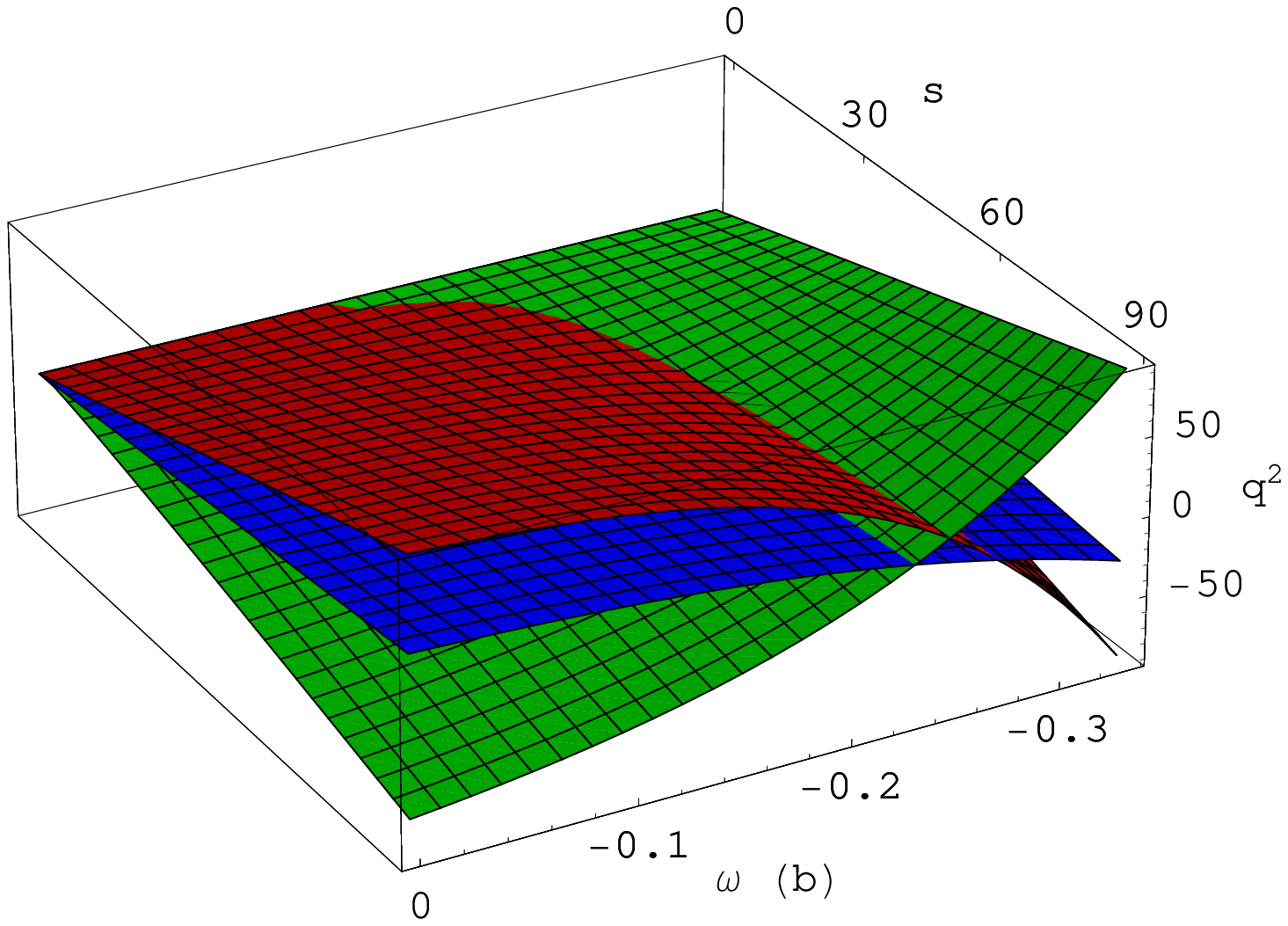}\\
  \caption{(a) Plots of $s_1\equiv s_1(\om)$ (red), $s_2\equiv s_2(\om)$ (blue) and $s_3\equiv s_3(\om)$ (green) for $-1/3<\om<0$ and $q=c=1$. $T>0$ for $s>s_1$, $C_q>0$ for $s_1<s<s_2$ and $M>0$ for $s>s_3$. The red, brown and cyan curves are the isotherms $T=0$, $T=1/(8\pi \sqrt{2})$ and $T=9/(4\pi 10^{3/2})$. The curves $s_2\equiv s_2(\om)$ (blue) and $s_3\equiv s_3(\om)$ (green) intersect at the point $(\om_0,s_0)=(-0.19,21.37)$. The \abh surrounded by \q is locally stable in the region between the green and blue curves for $\om_0<\om<0$ and unstable in the region above the blue curve for $\om_0<\om<0$. Limited to these values of $\om$, the blue curve determines the points where $C_q$ diverges, there the \abh undergoes a second order phase transition. (b) Plots of the surfaces $q^2=s+6c \omega  s^{W_-}$ (red), $q^2=(s/3)+2c\omega (2+3\omega )s^{W_-}$ (blue) and $q^2=-s+2c s^{W_-}$ (green) for $c=1$. $T>0$ below the red surface, $C_q>0$ below the red surface and above the blue one and $M>0$ above the green one.}\label{Fig2}
\end{figure}

Fig.~\ref{Fig2} (a) shows plots of the curves $s_1\equiv s_1(\om)$ (red), $s_2\equiv s_2(\om)$ (blue) and $s_3\equiv s_3(\om)$ (green) for $-1/3<\om<0$. We have chosen $q=c=1$ in such a way that $s_4(\om)<s_1(\om)$ so that there is no need to plot the curve $s_4\equiv s_4(\om)$. $T>0$ for $s>s_1$, $C_q>0$ for $s_1<s<s_2$ and $M>0$ for $s>s_3$. Three isotherm curves are shown in Fig.~\ref{Fig2} (a): $T=0$ (red), $T=1/(8\pi \sqrt{2})$ (brown) and $T=9/(4\pi 10^{3/2})$ (cyan). The plane region of local stability, identified by $T>0$, $C_q>0$ and $M>0$, is the region enclosed by the green curve, blue one and the $s$ axis for $\om_0<\om<0$ where $(\om_0,s_0)=(-0.19,21.37)$ is the intersection point of the curves $s_2\equiv s_2(\om)$ (blue) and $s_3\equiv s_3(\om)$ (green). For $q$ held constant,  $(\om_0,s_0)$ lies on the curve given by
\begin{equation*}
    s=3\frac{1-\omega  (3 \omega +2)}{1+3 \omega  (3 \omega +2)}q^2\qquad (\forall\,c)\,.
\end{equation*}
The region where the hole is unstable is above the blue curve for $\om_0<\om<0$. Since $C_q$ diverges on the blue curve, this latter determines a limit for second order phase transition.

The conditions in~\eqref{4.9} and~\eqref{4.8} are rewritten as $T>0:\;q^2<s+6c \omega  s^{W_-}$, $C_q>0:\;q^2>(s/3)+2c\omega (2+3\omega )s^{W_-}$ (with $T>0$) and $M>0:\;q^2>-s+2c s^{W_-}$, respectively. Fig.~\ref{Fig2} (b) shows plots of the surfaces $q^2=s+6c \omega  s^{W_-}$ (red), $q^2=(s/3)+2c\omega (2+3\omega )s^{W_-}$ (blue) and $q^2=-s+2c s^{W_-}$ (green) for $c=1$. The curves $s_1\equiv s_1(\om)$ (red), $s_2\equiv s_2(\om)$ (blue) and $s_3\equiv s_3(\om)$ (green) are projections of intersections of these surfaces with the plane $q^2=1$. The physical space region is all of the space region bounded above by the red surface and below by the green one if there $q^2>0$ or the subregion of it where $q^2>0$. If the point $p$ representing the thermodynamic state of the \abh lies between the red and the blue surfaces and above the green one ($M>0$), the hole is thermodynamically stable against fluctuations in ($s,q$): fluctuations are ``entropically'' suppressed. Since $T$ does not fluctuate, the stability condition is subject to the further constraint: r.h.s of~\eqref{3.4} $=$ constant. For fixed $c$, this defines a new surface in the space ($\om,s,q^2$) of parameters. Stability concerns only those states $p$ of the hole which lie on the segment of this new surface $T=\text{constant}$ which is sandwiched by the red and blue surfaces. The space region of instability is bounded above by the part of the blue surface where $q^2>0$ and below either by the green surface if there $q^2>0$ or by the plane $q^2=0$.

For $\om=-1/3$ ($0<c<1/2$) we have seen that the solution is an ordinary RN \aBH, given by~\eqref{2.4}, with new mass $M'$ and charge $q'$. Using~\eqref{3.6a} we obtain
\begin{equation}\label{4.13}
    C'_q=(1-2c) \frac{2\pi s(s- q'^2)}{3q'^2-s}\,.
\end{equation}
[Recall that in this case $s\equiv s'/(1-2c)=(M'+\sqrt{M'^2-q'^2})^2$]. Thus the conditions for local stability are the same as in~\eqref{4.10} but with $M'$ and $q'$, leading to
\begin{equation}\label{4.14}
    3/4<(1-2c)q^2/M^2<1\,.
\end{equation}
By~\eqref{3.00} we see that \bh with either $q^2\leq M^2$ or $q^2> M^2$ are locally stable provided conditions~\eqref{4.14} are satisfied. Black holes with $3M^2/4>(1-2c)q^2$ are unstable for $C'_q<0$. In the ($M^2,q^2$)-plane, the line $M^2=4(1-2c)q^2/3$ determines a phase transition of second order where $C'_q$ diverges.

\subsubsection{\pmb{$T$} fluctuating}\label{4a12}

In processes where $T$ fluctuates, the conditions in~\eqref{4.6} need be solved simultaneously. The second condition in~\eqref{4.6} is already satisfied; using~(\ref{3.3}, \ref{3.4}, \ref{4.7}), the third one leads to
\begin{equation}\label{4.15}
    \frac{q^2-s-6c\omega (2+3\omega )s^{W_-}}{3q^2-s-6c\omega (2+3\omega )s^{W_-}}>0\,.
\end{equation}
Note that if the numerator of the fraction in~\eqref{4.15} is positive, the denominator, which also appears in the expression of $C_q$, too will be positive. Thus, all we have to solve are the conditions~(\ref{4.8}, \ref{4.9} (a)) and the new one $q^2-s-6c\omega (2+3\omega )s^{W_-}>0$, all grouped into one system, respectively
\begin{align}
\label{4.16}&\;s+q^2>2c s^{W_-}\;(M>0)\\
\label{4.17}&\;s-q^2>-6c\omega s^{W_-}\;(T>0)\\
\label{4.18}&s-q^2<-6c\omega (2+3\omega )s^{W_-}\,.
\end{align}

We first consider an ordinary RN \abh ($c=0$). The last two conditions~(\ref{4.17}, \ref{4.18}) lead to $s=q^2$, then, by the first one, $s=q^2>0$. With $s=(M+\sqrt{M^2-q^2})^2$, we obtain $M=q$. Thus, only the extreme RN \abh is stable against fluctuations in ($s,q,T$). The first law takes the simple form $\dd M=\dd q$: for any change in $q$ there corresponds an equal change in $M$ so that the hole remains extreme for any $\dd q$.

In the generic case $c>0$, $-1/3<\om<0$ we have $1<2+3\om<2$. The solution of~\eqref{4.18} is similar to that of~\eqref{4.17}: the line $y(s)=s-q^2$ intersects the concave-down curve $y(s)=-6c\omega (2+3\omega )s^{W_-}$ at one and only one point $s_5(\om)>q^2$ [note that $s_5(\om)<s_2(\om)$]. Now, because $2+3\om>1$, the graph of $y(s)=-6c\omega (2+3\omega )s^{W_-}$ is above that of $y(s)=-6c\omega s^{W_-}$, the intersection of which with $y(s)=s-q^2$ determines the point $s_1(\om)>q^2$ as discussed earlier. This leads to $s_1(\om)<s_5(\om)$. Thus, the three conditions~(\ref{4.16} to \ref{4.18}) are solved by
\begin{equation}\label{4.19}
    s_1(\om)<s<s_5(\om)\;\text{ and }\;M>0\,.
\end{equation}

\begin{figure}[h]
\centering
  \includegraphics[width=0.47\textwidth]{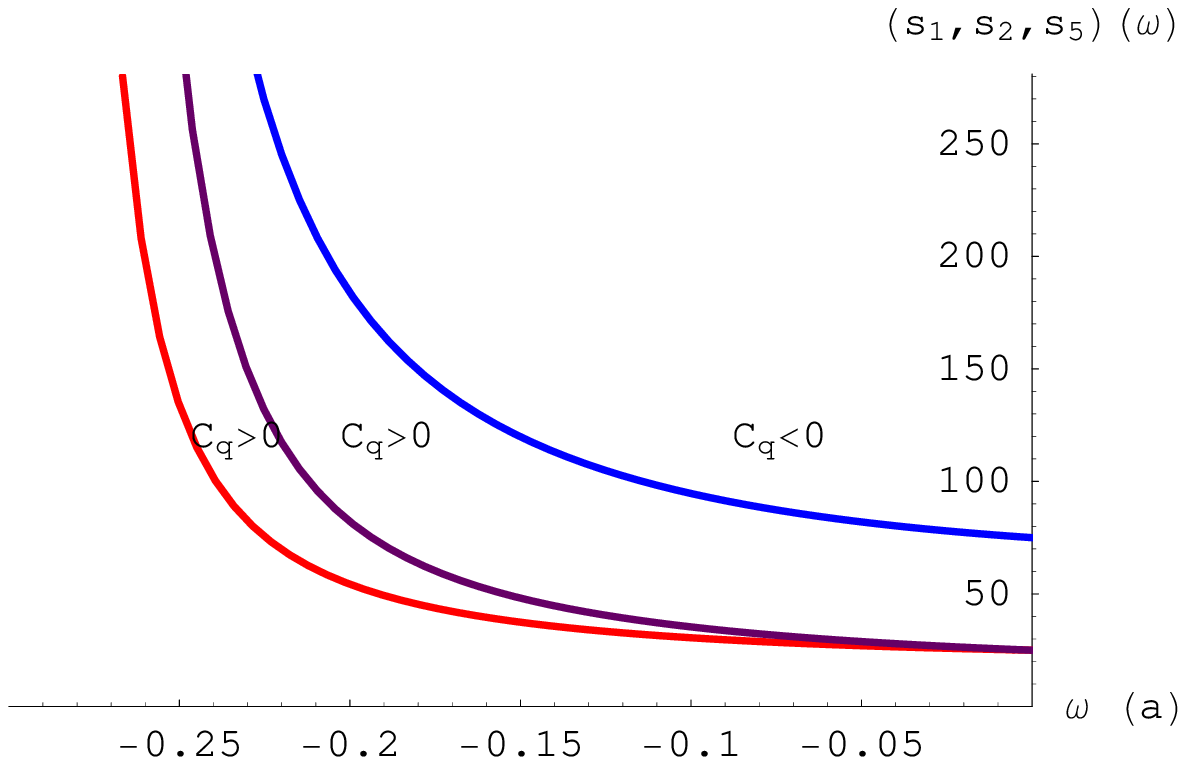}\includegraphics[width=0.47\textwidth]{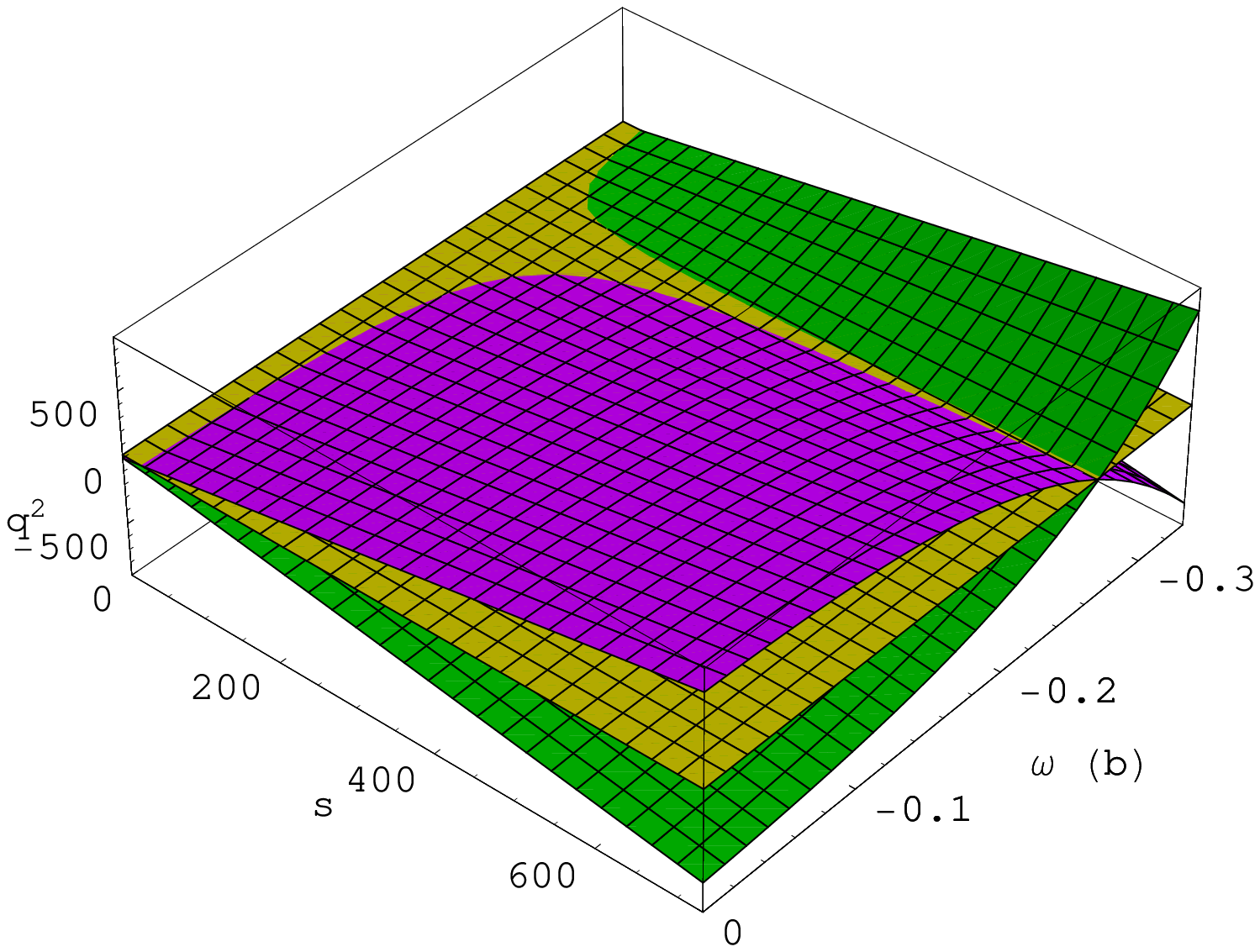} \includegraphics[width=0.47\textwidth]{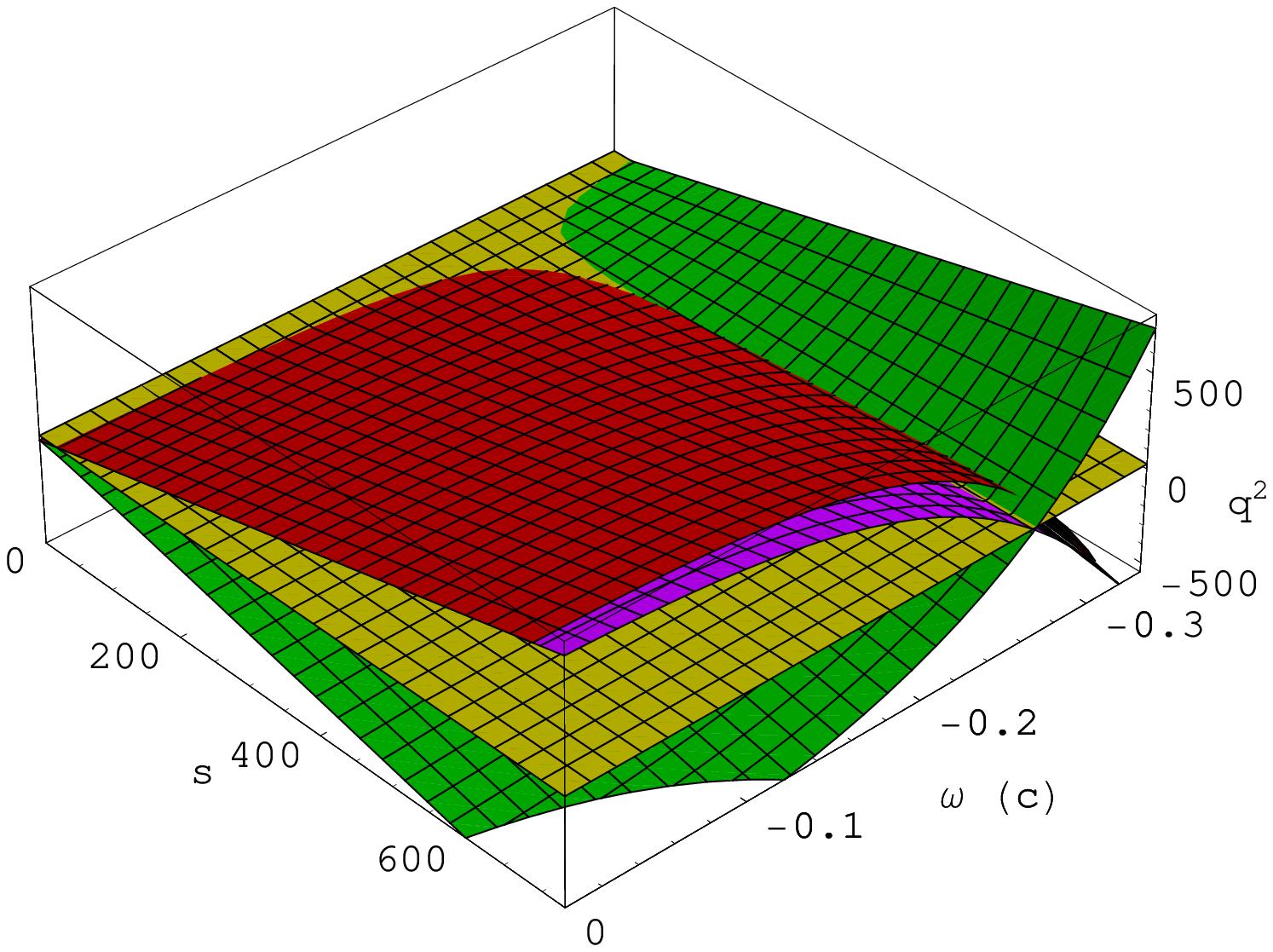}\\
  \caption{(a) Plots of $s_1\equiv s_1(\om)$ (red), $s_2\equiv s_2(\om)$ (blue) and $s_5\equiv s_5(\om)$ (magenta) for $-1/3<\om<0$, $c=1$ and $q=5$. (b) The space region of stability against fluctuations in both ($s,q,T$) is above the magenta surface and (c) below the red one and where the green surface is below both of them.}\label{Fig3}
\end{figure}

Allowing fluctuations in $T$, the condition for local stability has been narrowed since $s_5(\om)<s_2(\om)$ (compare with~\eqref{4.12}). This is the region bounded below by the red curve and above by the magenta one of Fig.~\ref{Fig3} (a), which shows plots of the curves $s_1\equiv s_1(\om)$ (red), $s_2\equiv s_2(\om)$ (blue) and $s_5\equiv s_5(\om)$ (magenta) for $-1/3<\om<0$, $c=1$ and $q=5$. The three curves do not intersect and admit the line $\om=-1/3$ as vertical asymptote. For fixed ($\om,q$), as the entropy increases from the red line which defines the states of extreme \BH, the state of the hole crosses the magenta line and becomes unstable regarding fluctuations in $T$. However, this transition cannot be qualified a first order phase transition since the entropy is continuous there (no jump in $s$) and the phase of the hole is the same; rather one might qualify it a ``behavioral'' change or transition. In the region bounded below by the magenta curve and above by the blue one, the hole is, however, stable regarding changes in $s,q$, as we have seen in the previous subsection. So this is a phase transition from thermodynamic states, which are stable against fluctuations in ($s,q,T$), to states which are stable against fluctuations in ($s,q$) only. If the entropy continues to increase, the hole undergoes the above-mentioned second order phase transition by crossing the blue line where $C_q$ diverges and changes the sign.

Fig.~\ref{Fig3} (b) and Fig.~\ref{Fig3} (c) show plots of the surfaces $q^2=s+6c \omega  s^{W_-}$ (red), $q^2=s+6c\omega (2+3\omega )s^{W_-}$ (magenta), $q^2=-s+2c s^{W_-}$ (green) and the plane $q^2=25$ (yellow) for $c=1$ [for clarity, the blue surface $q^2=(s/3)+2c\omega (2+3\omega )s^{W_-}$ is not shown]. The plane $q^2=25$ intersects the red surface, which represents the states of extreme \BH, along a curve the projection of which on the ($\om,s$)-plane is the red curve of Fig.~\ref{Fig3} (a). The other curves in Fig.~\ref{Fig3} (a) are also projections of intersections of the plane with the corresponding surfaces.
For fixed ($\om,s$), as $q^2$ decreases, along a vertical line [Fig.~\ref{Fig3} (c)], from its value on the red surface to its value on the magenta one, the hole remains stable against fluctuations in $T$ till the line crosses the magenta surface and the hole undergoes a behavioral change. As the charge decreases again, the state of the hole crosses the blue surface (not shown) and undergoes a second phase transition of second order.

We have thus shown that \bh with low entropy, or high charge or both are stable against fluctuations in ($s,q,T$), and \bh with relatively high entropy, or low charge or both are stable against fluctuations in ($s,q$) only. There is a behavioral change between the states of these holes maintaining the sign of $C_q>0$. Finally, \bh with high entropy, or low charge or both are unstable against allowed fluctuations and have $C_q<0$.

For $\om=-1/3$ ($0<c<1/2$), using Eqs.~(~\ref{3.7}, \ref{4.13}) we bring the third condition in~\eqref{4.6} to
\begin{equation}\label{4.20}
    \frac{q'^2-s}{3q'^2-s}>0\,.
\end{equation}
The problem is similar to that of ordinary RN \bh with $M$ and $q$ replaced by $M'$ and $q'$: Solving~\eqref{4.20} together with $T'>0$ and $C'_q>0$ we obtain $M'=q'$ or $M=\sqrt{1-2c}\,q$. Only the extreme \abh is stable against fluctuations in ($s,q$). The first law takes the simple form $\dd M=\sqrt{1-2c}\,\dd q$: for any change in $q$ there corresponds a proportional change in $M$ so that the hole remains extreme for any $\dd q$.

\subsection{ME and CE by the TPM}\label{4b}

We apply the TPM to the case of an isolated RN \abh surrounded by \Q, which is the ME and extend the analysis to the case where the hole is immersed in a heat bath, which is the CE treated in the preceding subsection by the classical thermodynamic approach. The aim of applying the TPM to the CE is to rederive in an elegant way the results of Sects.~\ref{4a11} and~\ref{4a12} using the powerful method of Poincar\'e~\cite{Poin}.

\begin{figure}[h]
\centering
  \includegraphics[width=0.35\textwidth]{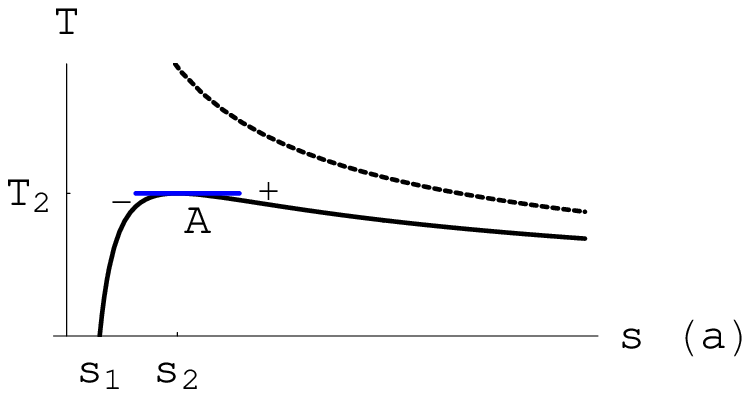} \includegraphics[width=0.35\textwidth]{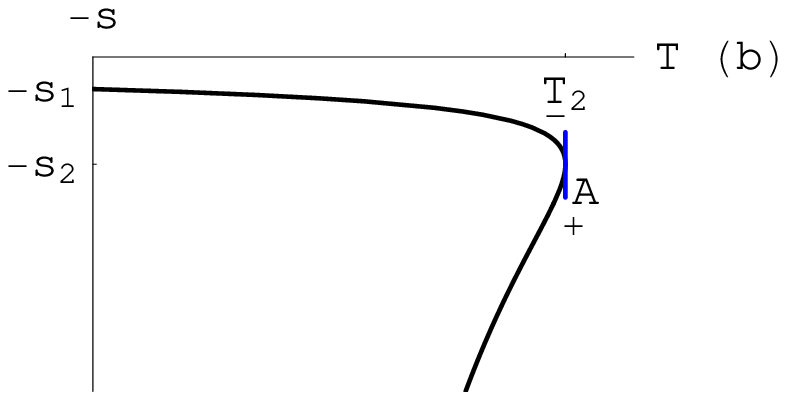} \includegraphics[width=0.35\textwidth]{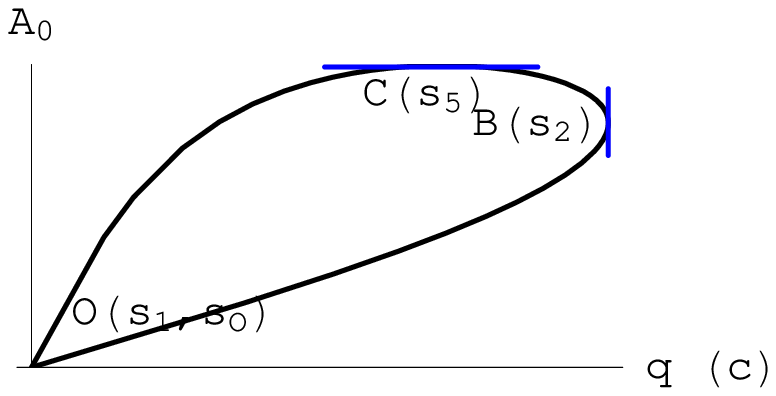}\\
  \caption{(a) $(q=1,c=1,\om=-0.1)$: Plot of $T$ vs. $s$ at constant $q$ for both the isolated (ME) RN \abh surrounded by \q (fitted line) and \s \abh (dotted line). (b) $(q=1,c=1,\om=-0.1)$: Plot of $-s$ vs. $T$ at constant $q$ for the RN \abh surrounded by \q immersed in a heat bath (CE). (c) $(T=0.05,c=1,\om=-0.1)$: Plot of $A_0$ vs. $q$ at constant $T$ for the RN \abh surrounded by \q immersed in a heat bath (CE). It is a parametric plot for $s_1\leq s\leq s_2$, $s_2\leq s\leq s_O$, $s_1=0.329134$, $s_2=7.191$, $s_O=14.65$, $q=\sqrt{s-4Ts^{3/2}+6c\om s^{W_-}}$ and $A_0=q/\sqrt{s}$.}\label{Fig4}
\end{figure}

In the TPM one uses the following Massieu functions $\Psi$: ($s,-F/T,-G/T,\cdots$) or, equivalently, the thermodynamic potentials $\Psi$: ($M,F,G,\cdots$)~\cite{tr1} depending on the ensemble, where $F=M-Ts$ and $G=F-A_0q$ are the Helmholtz and Gibbs free energies, respectively. For our ensembles the corresponding potentials at equilibrium are:
\begin{align}
\label{4.21}&\text{ME: }\Psi=M(s,q)\\
\label{4.22a}&\text{CE: }\Psi=F(T,q) \text{ if } T \text{ constant}\\
\label{4.22b}&\text{CE: }\Psi=G(T,A_0) \text{ if } T \text{ fluctuates}\,.
\end{align}
The TPM consists in plotting the planar curves $\partial\Psi/\partial x$ (all other variables kept constant) against $x$, which is some control parameter. The variable $\partial\Psi/\partial x$ is called the conjugate of $x$ with respect to $\Psi$. These curves are called linear series of equilibrium. Changes of equilibrium (from stable to less stable to unstable and conversely) occur at points where the curves have vertical tangents or bifurcations. If the curve has no vertical tangents, as in Fig.~\ref{Fig4} (a), then all points on the curve have the same degree of stability: If it is known that a point on the curve represents a stable equilibrium state, then all the points on the curve represent similar states. If the curve, with vertical tangent, is concave left near the vertical tangent [as shown in Fig.~\ref{4} (b,c)], then \textsl{all the points} on the branch of the curve where the slope is negative \textsl{near} the vertical tangent [upper branch in Fig.~\ref{Fig4} (b,c)] are more stable than the points on the other branch where the slope is positive \textsl{near} the vertical tangent\footnote{The slope need not be of the same sign along a given branch. The rule is valid if $\Psi$ is one of the potentials ($M,F,G\cdots$), which are minimum for stable equilibria. For $\Psi$ chosen from the list of Massieu functions ($s,-F/T,-G/T,\cdots$), which are maximum for stable equilibria, then if the curve is concave left, all the points on the branch of the curve where the slope is positive near the vertical tangent are more stable than the points on the other branch where the slope is negative near the vertical tangent.} [lower branch in Fig.~\ref{Fig4} (b,c)].

For the ME, the control parameter is the entropy $s$. Since $(\partial \Psi/\partial s)_q=(\partial M/\partial s)_q=T$ we have plotted in Fig.~\ref{Fig4} (a) the series of equilibria $T(s)$, with $q$ constant, for both the isolated RN \abh surrounded by \q (fitted line) and \s \abh (dotted line). The curves approach each other as $s\to\infty$. Knowing that the isolated \s \abh is locally stable~\cite{tr1}, we conclude that the isolated RN \abh surrounded by \q is at least stable for large values of $s$. But since there is no change of stability on the series of equilibrium [no vertical tangents on the curve $y=T(s)$], the hole is then stable for all $s$. Notice that this conclusion remains valid if we assume $\dd c\neq 0$ since we could take $\Psi=M(s,q,c)$.

The case of the CE has been split into tow subcase: $T$ constant (subsubsection~\ref{4a11}) and $T$ fluctuating (subsubsection~\ref{4a12}). In the former case, $\Psi=F(T,q)$ and $T$ is a control parameter. Using the first law, $\dd M=T\dd s+A_0\dd q$ (with $\dd c= 0$), we obtain
\begin{equation}\label{4.23}
    \dd F=-s\dd T+A_0\dd q\,.
\end{equation}
Hence, $(\partial \Psi/\partial T)_q=(\partial F/\partial T)_q=-s$. We need to plot the curve $-s(T)$ for $q$ constant. This is the same curve plotted in Fig.~\ref{Fig4} (a) but rotated $90^{\circ}$ clockwise as shown in Fig.~\ref{Fig4} (b). Notice that the point $A$ where $-s(T)$ has a vertical tangent is the same point where $T(s)$ has a horizontal tangent. Using~\eqref{3.4}
\begin{equation}\label{4.24}
    \Big(\frac{\partial T}{\partial s}\Big)_{q,c}=\frac{3 q^2-s-6 c \omega  (2+3 \omega ) s^{W_-}}{8 s^{5/2}}
\end{equation}
we see that $(\partial T/\partial s)_{q,c}=0$ at the points $s_2$ where $C_q$ diverges. Thus we reach the same conclusion as in Sect.~\ref{4a11}: For fixed ($q,c,\om$), the stability breaks at the point $A$ where there is a vertical tangent (where $C_q=\infty$), the points where $T<T_2=T(s_2)$ and $s_1<s<s_2$ are stable (negative slopes) and the points where $T<T_2$ and $s>s_2$ are unstable (positive slopes).

For the ordinary RN \abh ($c=0$), $s_1=q^2$, $s_2=3q^2$ and $T_2=1/(6\sqrt{3}q)$. This is the maximum temperature beyond which the hole cannot resist evaporation.

Using~\eqref{4.23}, we have $(\partial \Psi/\partial q)_T=(\partial F/\partial q)_T=A_0$. So we could also plot $A_0(q)$ for $T$ constant, using thus $q$ as a control parameter. This is a parametric plot, shown in Fig.~\ref{Fig4} (c), parameterized by $s$. The entropy increases along the upper branch (stable) from $s_1$ at $O$ to $s_2$ at $B$, it continues to increase along the lower branch (unstable) from $B$ back to $O$. To show that, at $B$, $s=s_2$, which is the point where $C_q$ diverges, we rewrite~\eqref{3.4} as $q=\sqrt{s-4Ts^{3/2}+6c\om s^{W_-}}$. With $T$ constant ($(c,\om)$ are also kept constant), at $B$ we have $\dd q/\dd s=0$ leading to $s-6Ts^{3/2}+6c\om W_-s^{W_-}=0$. Eliminating $T$ by~\eqref{3.4}, the remaining equation is again $3 q^2-s-6 c \omega  (2+3 \omega ) s^{W_-}=0$ [compare with~\eqref{4.24}], the solution of which is $s=s_2$. Here again we reach the same conclusion as in subsubsection~\ref{4a11}.

If, in the CE, we assume that $T$ fluctuates, the appropriate potential is the Gibbs function $\Psi=G(T,A_0)$ leading to $\dd G=-s\dd T-q\dd A_0$. Using $A_0$ as a control parameter, we obtain $(\partial \Psi/\partial A_0)_T=(\partial G/\partial A_0)_T=-q$. So we need to plot the curve $-q(A_0)$ for $T$ constant. This is the same curve plotted in Fig.~\ref{Fig4} (c) but rotated $90^{\circ}$ clockwise. So the horizontal tangent at $C$ becomes a vertical one and the (upper) branch from $s_1$ at $O$ to $s_5$ at $C$ is the stable one if fluctuations in $T$ are taken into consideration [negative slopes near $C$ after revolving the curve in Fig.~\ref{Fig4} (c) $90^{\circ}$ clockwise]. The branch from $C$ through $B$ back to $O$ is the unstable one regarding fluctuations in $T$. So we just re-derived the conclusion made in subsubsection~\ref{4a12} (compare with~\eqref{4.19}). All we need to show is $s=s_5$ at $C$. At $C$, $\dd A_0/\dd s=0$ which is the same as $2Ts^{3/2}+3c\om (1+3\om)s^{W_-}=0$. Eliminating $T$ by~\eqref{3.4}, the remaining equation is again $q^2-s-6 c \omega  (2+3 \omega ) s^{W_-}=0$ [compare with~\eqref{4.18}], the solution of which is $s=s_5$.

\section{\large Phase transition via geometric methods}\label{5}

In this section we briefly present the results of thermodynamic stability as derived by the two geometric approaches: (1) Geometrothermodynamics (GTD)~\cite{quevedo1} and (2) Liu-Lu-Luo-Shao (LLLS) method~\cite{liu}.

\subsection{Geometrothermodynamics}

Let $\Psi$ be a thermodynamic potential and ($E^a,I^a$) the set of associated extensive and intensive variables. We define a $(2n+1)$-dimensional space $\mathbb{T}$ whose coordinates is the set $Z^A=\{\Psi,E^a,I^a\}$ where $A: 0\to 2n$ and $a:1 \to n$. Together with the Gibbs 1-form $\Ta =\dd \Psi-\de_{ab}I^a\dd E^b$, ($\mathbb{T},\Ta$) make up the $(2n+1)$-dimensional contact manifold of metric $G^{AB}(Z^C)$, which is the thermodynamic phase space~\cite{quevedo1}. Here we assume that $\mathbb{T}$ is differentiable and that $\Ta$ satisfies the condition $\Ta\wedge (\dd \Ta)^n\neq 0$. The subspace $\mathbb{E}\subset \mathbb{T}$ of equilibrium states is defined by the map: $\varphi: \mathbb{E}\to\mathbb{T}$ such that the pullback vanishes at $\Ta$: $\varphi^*(\Ta)\equiv 0$.

The singularities of the curvature scalar $R_{GTD}$ of $\mathbb{E}$ determine the points or states where there are second order phase transitions of the thermodynamic system. The metric of the space $\mathbb{E}$ is given by~\cite{manuel}
\begin{eqnarray}
\dd l^{2}_{\text{GTD}}=\left(E^{c}\frac{\partial\Psi}{\partial E^{c}}\right)\left(\eta_{ad}\delta^{di}\frac{\partial^{2}\Psi}{\partial E^{i}\partial E^{b}}\right)\dd E^{a}\dd E^{b}\label{5.1}\;,
\end{eqnarray}
where $\eta_{ad}=(-1,1,...,1)$. This metric is invariant under Legendre transformations~\cite{quevedo1}.

We consider an RN \abh surrounded by \q and immersed in a heat bath at fixed temperature, we assume no fluctuations in $c$ ($\dd c=0$): This is the CE subject to~\eqref{4.4}. It is then convenient to use the mass (\ref{3.1}) as the thermodynamic potential. We can write the matrix (\ref{5.1}) in a metric form
\begin{eqnarray}
\dd l^{2}_{\text{GTD}}&=&\left(s\frac{\partial M}{\partial s}+q\frac{\partial M}{\partial q}\right)\left(-\frac{\partial^{2} M}{\partial s^2}\dd s^{2}+\frac{\partial^{2} M}{\partial q^2}\dd q^{2}\right)\nonumber\\
\label{5.2}&=&\frac{Y}{8s^{5/2}}\{[s-3q^2+6c\omega (2+3\omega)s^{W_-}]\dd s^2+8s^2\dd q^2\}
\end{eqnarray}
and the associated curvature scalar reads
\begin{multline}\label{5.3}
R_{\text{GTD}}=-\frac{X}{Y^3} \Big[\frac{(-3 q^2+s-18 c \omega ^2 s^{W_-})^2+18 q^2 [-3 q^2 +s+6 c \omega  (2+3 \omega ) s^{W_-}]}{64 s^3}\Big]\\+\frac{X^2}{2Y} \Big[\frac{2 s+3 c \omega  (10+21 \omega +9 \omega ^2) s^{W_-}}{16 s^{9/2}}\Big]+\frac{3}{2 Y^2} \Big\{X \,\frac{q^2- s+6 c \omega ^2 (5+6 \omega ) s^{W_-}}{16 s^{5/2}}-1\\
+X^2\,\frac{(3 q^2-s+18c \omega ^2 s^{W_-}) [-5 q^2+ s+2 c \omega  (8+18 \omega +9 \omega ^2) s^{W_-}]}{128 s^5}\Big\}
\end{multline}
where
\begin{align}
\label{5.4}&X=\frac{C_q}{T}=\frac{8s^{5/2}}{3q^2-s-6c\om(2+3\om)s^{W_-}}\\
\label{5.5}&Y=sT+qA_0=\frac{3q^2+s+6c\om s^{W_-}}{4\sqrt{s}}\,.
\end{align}

It is obvious that $R_{\text{GTD}}$ diverges at the points where $X$ diverges corresponding to $C_q=\infty$. A divergence in the value of $C_q$ announces a second order phase transition as derived in subsection~\ref{4a11}. $R_{\text{GTD}}$ diverges at $Y=0$ too but this equation, based on the analysis made in Sects.~\ref{4a11} and~\ref{4a12},  represents no physical effect. This is a pathologic effect attributable to the fact that the metric~\eqref{5.2} becomes singular at $Y=0$.

\subsection{Liu-Lu-Luo-Shao method}

The second geometric method was developed later by Liu-Lu-Luo-Shao (LLLS)~\cite{liu}. The method relies on the same principal as the first one in that any singularity in the curvature scalar of the associated metric signals a second order phase transition of the system. The metric of this space is simply the Hessian matrix of the Helmholtz free energy which, in the CE, is given by
\begin{eqnarray}
\dd l^{2}_{\text{LLLS}}=-\dd T\dd s+\dd A_{0}\dd q=-\frac{\partial T}{\partial s}\dd s^2+\left(\frac{\partial A_{0}}{\partial s}-\frac{\partial T}{\partial q}\right)\dd s\dd q+\frac{\partial A_{0}}{\partial q}\dd q^2\label{5.5}\,.
\end{eqnarray}
Using the expressions~(\ref{3.3}) and~(\ref{3.4}), (\ref{5.5}) becomes
\begin{equation}\label{5.6}
    \dd l^{2}_{\text{LLLS}}=\frac{1}{8s^{5/2}}\{[s-3q^2+6c\omega (2+3\omega)s^{W_-}]\dd s^2+8s^2\dd q^2\}\qquad (\dd l^{2}_{\text{GTD}}=Y\dd l^{2}_{\text{LLLS}})\,.
\end{equation}
This metric provides the following curvature scalar
\begin{equation}\label{5.7}
R_{\text{LLLS}}=\frac{[2s+3c\omega (10+21\omega+9\omega^2)s^{W_-}]}{32s^4}\,X^2\,.
\end{equation}
Here again we clearly see that the divergence of the curvature scalar (\ref{5.7}) corresponds to that of $C_q$, thus by the LLLS  method we are too able to locate the points where the second order phase transition takes place.

\section{\large Conclusion}\label{6}

We have seen that asymptotically flat RN \bh surrounded by \q have higher entropies than ordinary ones. The excess in entropy is attributable to the entropy of \q matter. For a given value of the \q density $c$, they commulate more electric charges than ordinary RN holes before they become naked singularities. For $c\ll 1$, the maximum relative cumulated charge $(q^2-M^2)/M^2$ is proportional to $c/M^{3\om+1}$, if $-1/3<\om<0$, or to $2c/(1-2c)$, if $\om=-1/3$.

Taking $c$ as a thermodynamic variable, as some works have done for the cosmological constant~\cite{cosmo1,cosmo2} and for $c$ also~\cite{cit1}, we have obtained the generalized Smarr formula and the first law of thermodynamics.

As one charges adiabatically an ordinary RN \aBH, it cumulates mass and charge till their total values equate the radius of the horizon, which remains constant during the process. If the RN \abh is surrounded by \q then it will cumulate more mass and charge, however, with their totals never exceeding the radius of the horizon.

Applying the classical thermodynamic method and restricting ourselves to the CE we have obtained generalized conditions for local stability of RN \bh surrounded by \Q. These conditions are the shifting of the known ones for ordinary RN \bh if fluctuations in $T$ are not allowed; If $T$ fluctuates, the same conditions apply with their upper limits constrained to lower values. We have reached the conclusion that, allowing fluctuations in $(s,q,T)$, only \bh with low entropy, or high charge or both are stable, while \bh with relatively high entropy, or low charge or both are stable against fluctuations in $(s,q)$ only. Between these and the other \abh states there is a behavioral change maintaining the sign of $C_q$.

We have also shown that the CE is unstable if all thermodynamic variables are allowed to vary.

We have completed the analysis of stability by applying the TPM and obtained an upper limit for the temperature [$T_2=T(s_2)$] beyond which the CE is no longer stable, if fluctuations in $T$ are not allowed. For ordinary RN \bh $T_2=1/(6\sqrt{3}q)$. If $T$ fluctuates, we have obtained an upper limit for the electric potential on the horizon [$A_0(s_5)$] beyond which the CE is no longer stable. Another general result we could derive is that isolated \BH, which constitute MEs, are stable.

By the two geometrical method (GTD, LLLS) we could also determine the states corresponding to a second order phase transition.

The stability of non-asymptotically flat solutions is more involved and constitutes the matter of a subsequent work along with the case where $c$ fluctuates.

\section*{\large Acknowledgments}

M.~E.~Rodrigues thanks the UFES for the hospitality he has enjoyed during the development of this work and thanks the CNPq for partial financial support.

\section*{Appendix: Hessian analysis of the general case ($\dd s\neq0,\dd q\neq0,\dd c\neq0$)}
\renewcommand{\theequation}{A.\arabic{equation}}
\setcounter{equation}{0}

In this section we extend the work done in subsection~\ref{4a}, that is we will only discuss the CE case within the context of the classical thermodynamic approach allowing $c$ to vary. We consider the general case, ($\dd s\neq0,\dd q\neq0,\dd c\neq0$), of a \abh immersed in a thermal bath at constant temperature. Using~\eqref{4.5} along with $(\partial^2 M/\partial s \partial c)_{q}=(\partial T/\partial c)_{s,q}$ and the fact that $(\partial^2 M/\partial c^2)_{s,q}=(\partial^2 M/\partial q \partial c)_{s}\equiv 0$, which is derived from~\eqref{3.1}, we arrive at the following Hessian equilibrium condition that generalizes~\eqref{4.4}
\begin{equation}\label{A1}
    (\dd s,\dd q,\dd c)\underbrace{\begin{pmatrix}
T/C_q & (\partial T/\partial q)_{s,c} & (\partial T/\partial c)_{s,q} \\
 (\partial T/\partial q)_{s,c} & (\partial A_0/\partial q)_{s,c} & 0 \\
 (\partial T/\partial c)_{s,q} & 0 & 0
\end{pmatrix}}_H\begin{pmatrix}
\dd s \\
\dd q \\
\dd c
\end{pmatrix}>0,\,\text{for all }(\dd s,\dd q,\dd c).
\end{equation}
[The Hessian condition~\eqref{4.4} is represented by the upper left 2$\times$2 block of the above 3$\times$3 $H$ matrix]. To satisfy the equilibrium condition~\eqref{A1}, all the eigenvalues ($p_1,p_2,p_3$) of the 3$\times$3 $H$ matrix should be positive. The characteristic polynomial of $H$ is brought to the form
\begin{multline}\label{A2}
    C_qp^3-[T+C_q(\partial A_0/\partial q)_{s,c}]p^2\\+\{T(\partial A_0/\partial q)_{s,c}-C_q[(\partial T/\partial q)_{s,c}^2+(\partial T/\partial c)_{s,q}^2]\}p+C_q(\partial A_0/\partial q)_{s,c}(\partial T/\partial c)_{s,q}^2.
\end{multline}
Now, by~\eqref{3.3} we have $(\partial A_0/\partial q)_{s,c}=1/\sqrt{s}>0$, thus in~\eqref{A2} the coefficient of $p^3$, $C_q$, and the independent term, $C_q(\partial A_0/\partial q)_{s,c}(\partial T/\partial c)_{s,q}^2$, have the same sign. Consequently, the three eigenvalues ($p_1,p_2,p_3$) can't \emph{all} be positive and this implies that the CE is unstable if $c$ is allowed to vary.


\end{document}